\documentclass[twocolumn,showpacs,preprintnumbers,amsmath,amssymb]{revtex4}
\usepackage{graphicx}% Include figure files
\usepackage{dcolumn}% Align table columns on decimal point
\usepackage{bm}% bold math

\begin{document}

%\preprint{}


\title{Generic degeneracy and entropy in loop quantum gravity}

\author{Mohammad H. Ansari}
\email{mansari@perimeterinstitute.ca}
\affiliation{University of Waterloo, Waterloo, On, Canada N2L 3G1 }%
\affiliation{Perimeter Institute, Waterloo, On, Canada N2L 2Y5}%
\date{\today}% It is always \today, today,
\newcommand{\beq}{\begin{equation}}
\newcommand{\eeq}{\end{equation}}
\newcommand{\barr}{\begin{array}}
\newcommand{\earr}{\end{array}}
\newcommand{\ssz}{\scriptsize}
\newcommand{\amin}{a_{\mathrm{min}}}
\newcommand{\zmin}{\zeta_{\mathrm{min}}}

\begin{abstract}
Without imposing the trapping boundary conditions and only from
within the very definition of area it is shown that the loop
quantization of area manifests an unexpected degeneracy in area
eigenvalues. This could lead to a deeper understanding of the
microscopic description of a quantum black hole. If a certain number
of semi-classically expected properties of black holes are imposed
on a quantum surface its entropy coincides with the
Bekenstein-Hawking entropy.
\end{abstract}


%\preprint{}

\maketitle

\tableofcontents

\section{Introduction}

In the early works on black holes in loop quantum gravity
\cite{carloblackhole}, it was first understood that the black hole
entropy can be derived from the internal boundary of space without
imposing any boundary condition. The underlying details of this
pictures was later on recovered when the definition of a marginally
trapped surface \cite{kuchar} was rewritten in the Ashtekar-Sen
variables and this definition was extended into a quantum sector
\cite{Ashtekar:PhysRevLett}. Such a surface after quantization
contains a finite number of degrees of freedom and consequently
carries an entropy that coincides with the Bekenstein-Hawking black
hole entropy, \footnote{Recently a different approach towards this
entropy from the use of quantum information theory techniques in
loop quantum gravity has also been developed in
\cite{Livine:2005mw}.}. In this note by the use of the same setting
similar to the one of original works by Rovelli
\cite{carloblackhole}, we show that area operator acting on a
typical inhomogeneous surface state manifests a degeneracy that here
is worked out in both SO(3) and SU(2) group representations. This
degeneracy exhibits a scale invariant correlation with area with the
same exponent in both groups. However, this is not the only
degeneracy an area eigenvalue is left with. Recently it was
understood in \cite{Ansari:2006vg} that the complete spectrum of
area can be re-classified into different equidistant subsets. This
symmetry (the so-called ladder symmetry) increases the total
degeneracy such that the degeneracy of a large area eigenvalues
becomes proportional to its area exponentially. Moreover, we present
the derivation of the Bekenstein-Hawking expression for the entropy
of a Schwarzschild black hole of large surface area by the use of
Dreyer's conjecture \cite{Dreyer:2002vy} that the minimal area of a
hole in its dominant configuration should be identified with the
emissive quanta from a perturbed black hole in highly damping mode.
This motivates a rather different picture of a quantum horizon whose
precise dynamical definition perhaps should be looked at from within
a spin foam model, \cite{Perez:2003vx}.

In loop quantization approach to quantum gravity the kinematical
state space is taken to be $L^2$(connections on SU(2) bundle). For
any graph with finitely many edges and vertices embedded in a
spatial manifold, the space of connections is $\mathrm{SU(2)}^n$ (or
alternatively $\mathrm{SO(3)}^n$) where $n$ is the number of edges.
In fact, a connection on a graph tells us how to parallel transport
information along each edge of that graph. The canonical conjugate
of this connection field represents the quantum geometry of space.
The space of physical states is obtained by imposing constraints:
the gauge-invariance, the diffeomorphism invariance, and the
invariance under time evolution.

Given a graph and a surface in space, the area operator is supposed
to be the quantum analog of the usual classical formula for the area
of $S$. This operator only cares about the points where the graph
intersects the surface. A subset of surface states which has no node
residing on the surface was originally considered by Smolin and
Rovelli where they derived their eigenvalues in
\cite{Smolin:1994ge}. Later on, all possible states of a surface
were considered and the complete spectrum of area eigenvalues were
found from different approaches, \cite{ashtekarea}. From the
calculation, it was uncovered that those edges which are completely
tangential to the surface do not contribute in the surface area,
albeit the tangent vectors of crossing edges do contribute in it.

An edge with respect to an underlying surface falls into two
classes. It may

\begin{itemize}
  \item cross the surface at one intersecting point and bends at the surface on the
point to induce a tangent vector on the surface, or
  \item reside completely tangential to the surface.
\end{itemize}

\begin{figure}[h]
\begin{center}
\includegraphics[width=5cm]{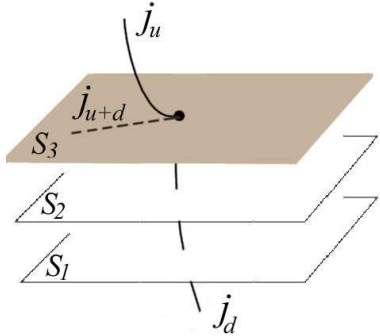}%
\end{center}
\caption{The position of a surface relative to spin network.}
\label{cap. classes of areas}
\end{figure}

A spin network with respect to three different surfaces $S_1$,
$S_2$, and $S_3$ was shown in Figure (\ref{cap. classes of areas}).
The quantum state of surface $S_3$ contains the bulk edges $j_u$ and
$j_d$ and their tangent vector $j_{u+d}$ at their joint node. Note
that the quantum states corresponding to the surfaces $S_1$ and
$S_2$ contain the edges of equal spins on both sides without bending
at these surfaces.

The area eigenvalue associated with a typical quantum state $\langle
j_u, j_{u+d}, j_d, j_t|$ is $a =  m_{j_u, j_{u+d}, j_d}\ a_o$,

\beq \label{eq. A degenerate sec.} m_{j_u, j_{u+d}, j_d} =
\sqrt{2f(j_u) + 2f(j_d) - f(j_{u+d})}, \eeq

where $f(x)=x(x+1)$ and $a_o := 4 \pi \gamma \ell_\mathrm{P}^2$. The
parameter $\gamma$ is a dimensionless parameter called the
Barbero-Immirzi parameter \cite{Immirzi:1996di} and
$\ell_{\mathrm{P}}$ is the Planck's length $\sqrt{\hbar G/ c^3}$.
Moreover, the tangent vector accept a finite number of quantum
values from the following spectrum:

\begin{equation}
\label{eq. j_u+d} j_{u+d} \in \{ |j_u - j_d|, |j_u - j_d|+1, \ldots,
j_u+j_d \} .
\end{equation}

\begin{figure}[h]
\begin{center}
\includegraphics[width=7cm]{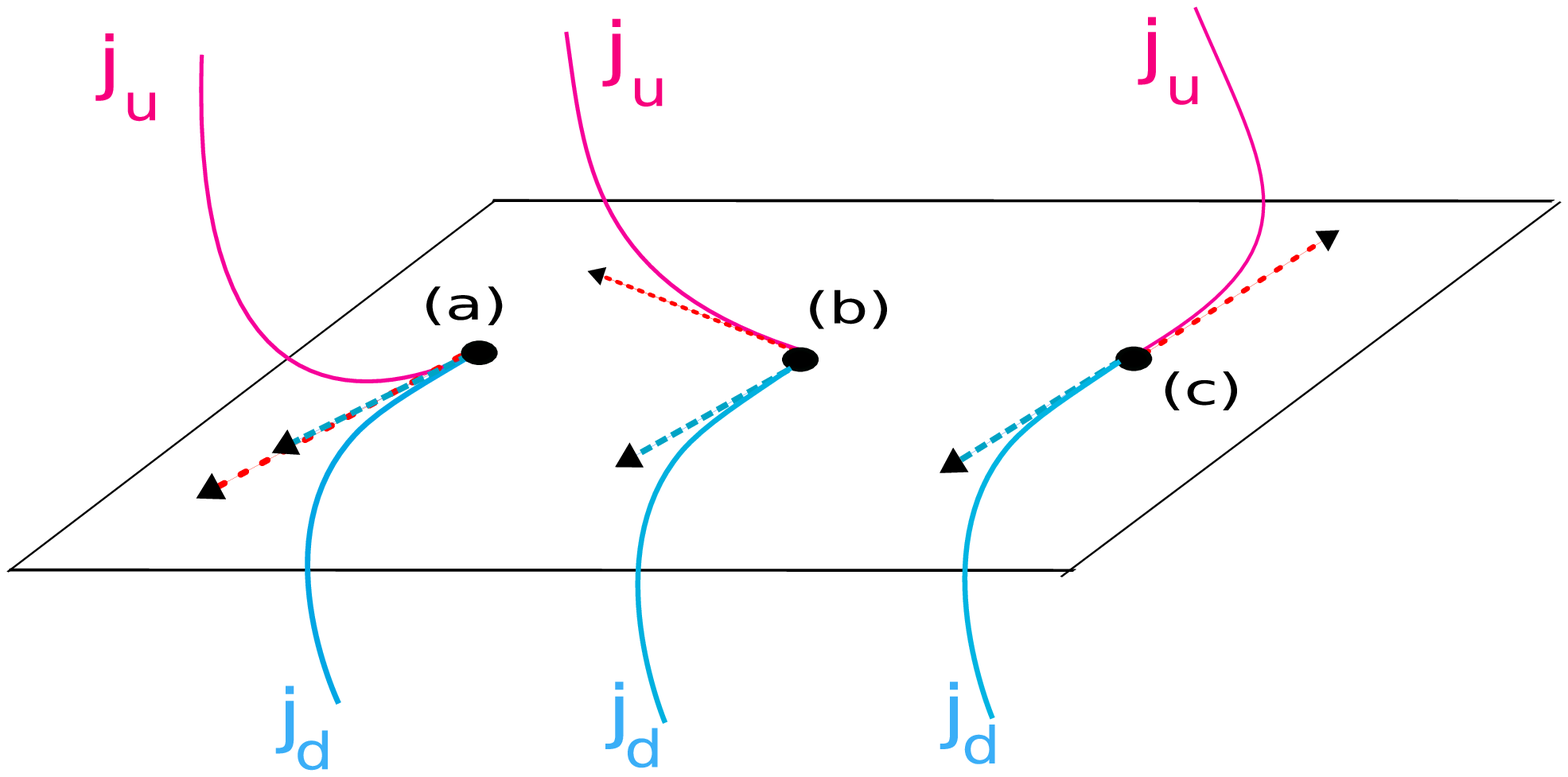}%
\end{center}
\caption{Different components of the tangent vector of the upper and
lower spins on the surface. The dashed arrows in blue and red colors
corresponds to the edges in the lower and upper side}\label{fig.jud}
\end{figure}

The tangent vector at a node is a spin between the sum and
difference of crossing edges at the node and resembles the total
vector of two quantum angular momenta (i.e. spin and orbital angular
momenta) in the Hydrogen atom. This pictorially is shown in Fig.
(\ref{fig.jud}) where the case (a) indicates $j_{u+d} = j_u + j_d$
and the case (c) indicates $j_{u+d} = |j_u - j_d|$. The case (b) in
Figure (\ref{fig.jud}) shows an intermediate value for the tangent
vector between the maximum and minimum.

Let us now describe the subset of state that was reported first by
Smolin and Rovelli as the area eigenstates in \cite{Smolin:1994ge}.
At a vertex if neither one of the upper and lower edges bends at the
surface, the tangent vector becomes zero. In the lack of the
completely tangential edges at the vertex, due to the gauge
invariance, the upper and lower spins must be equal. The quantum of
area in this case becomes $a = 2a_o \sqrt{j (j+1)}$. Therefore, the
subset includes only those states with puncturing edges without
bending at it.

Consider now a closed surface that divides space into two disjoint
subsets of interior and exterior regions.  Since such a surface has
no boundary a few additional vertices are needed in order to close
its corresponding spin network. This makes unwanted contributions to
the action of the area operator via the constraints:
\begin{equation}\label{eq. restrictions}
\barr{cc}
  \sum_{\alpha} j_u^{(\alpha)} \in \mathrm{Z}^+, &\ \   \sum_{\alpha} j_d^{(\alpha)} \in
  \mathrm{Z}^+,
  \earr
\end{equation}
where $\alpha$ indicates the vertices of the graph. Note that these
constraints are trivially satisfied in the SO(3) representation of
the spin network states, because the spins are already in pairs
(integer numbers). However, in SU(2) representation some spin
network states are excluded by (\ref{eq. restrictions}).

\section{Generic degeneracy}

For a given $j_u$ and $j_d$ a set of eigenvalues is generated from a
minimum  value (where $j_{u+d}=j_u+j_d$) to a maximum value (where
$j_{u+d}=|j_u - j_d|$) by (\ref{eq. A degenerate sec.}). Changing
either $j_u$ or $j_d$, a different finite set of area eigenvalues is
generated whose elements may or may not coincide with the elements
of other set of area eigenvalues. Generating area eigenvalues from
different eigenstates indicates that the spectrum of eigenvalues
becomes denser in larger area. All of the eigenvalues are
unexpectedly degenerate. In fact \emph{ there exist a finite set of
eigenstates that correspond to every area eigenvalue.}

%\subsection{Degeneracy in complete spectrum}

Let us consider the first four area eigenvalues in SO(3) group as
samples. These are $a_\mathrm{min} = \sqrt{2} a_o$, $a_2 = 2 a_o$,
$a_3 = \sqrt{6} a_o$, and $a_4 = 2\sqrt{2} a_o$, respectively. Note
this $a_4$ is the minimal eigenvalue of the subset that was first
discovered as the spectrum of area eigenvalues by integer spins.
This area has been considered many times in the literature as the
minimal area cell in the dominant configuration of a black hole
horizon (i.g. see
\cite{{carloblackhole},{Ashtekar:PhysRevLett},{Livine:2005mw},{Dreyer:2002vy}});
however it is the double of the the minimal area eigenvalue
$a_\mathrm{min}$ in the complete spectrum of (\ref{eq. A degenerate
sec.}).

The table (\ref{tab_so3}) shows the detail of area states
corresponding to these four area eigenvalues.

\begin{table}[h]  \centering
\begin{tabular}{|c|c|c|c||c|c|c|c||c|c|c|c|}  \hline
  % after \\: \hline or \cline{col1-col2} \cline{col3-col4} ..
   \ area\ & $j_u$ & $j_{u+d}$ & $j_d$ &\ area\ & $j_u$ & $j_{u+d}$ & $j_d$ &\ area\ & $j_u$ & $j_{u+d}$ & $j_d$  \\ \hline
  $a_\mathrm{min}$ & 0 & 1 & 1 & $a_3$ & 0 & 2 & 2 & $a_4$ & 1 & 0 & 1
  \\ \hline
  $a_\mathrm{min}$ & $1^{}$ & 1 & 0 & $a_3$ & 1 & 1 & 1 & $a_4$ & 1 & 4 & 3
  \\ \hline
  $a_\mathrm{min}$ & $1^{}$ & 2 & 1 & $a_3$ & 2 & 2 & 0 & $a_4$ & 3 & 4 & 1
  \\ \hline
  $a_2$ & 2 & 4 & $2^{}$ & $a_3$ & 2 & 5 & 3 & $a_4$ & 3 & 7 & 4
  \\ \hline
  $a_2$ & 2 & 3 & $1^{}$ & $a_3$ & 3 & 5 & 2 & $a_4$ & 4 & 7 & 3 \\
  \hline
  $a_2$ & 1 & 3 & $2^{}$ & $a_3$ & 3 & 6 & 3 & $a_4$ & 4 & 8 & 4 \\
  \hline
\end{tabular}
  \caption{The eigenstates corresponding to the first four SO(3) eigenvalues.}
  \label{tab_so3}
\end{table}

The three states corresponding to the minimal area are: 1) the state
with the upper spin one edge crossing the surface and bending at it
on the point. Such an edge induces all of its spin to the surface at
the point. However a completely tangential edge is necessary to lie
on the surface and ends at that point in order to make this state
gauge invariant off the surface. 2) The state with lower spin one
edge similar to the previous state. 3) The state with the upper and
lower spins one edges bending at the surface such that their tangent
vector come along each other in the same direction. This tangent
vector could also connect to other completely tangential excitations
on the surface.

In SU(2) group the eigenstates of the first six eigenvalues are
 tabulated in the table (\ref{tab_su2}). These eigenvalues are: $a_\mathrm{min} =
\frac{\sqrt{3}}{2} a_o $, $a_2 = a_o$, $a_3 = \frac{\sqrt{7}}{2}
a_o$, $a_4 = \sqrt{2} a_o$, $a_5 = \frac{\sqrt{11}}{2} a_o$, and
$a_6 = \sqrt{3} a_o$. Note that the sixth eigenvalue is in fact the
minimal area applied in the literature so far for the purpose of
black hole entropy calculation in this group.

\begin{table}[h]  \centering
\begin{tabular}{|c|c|c|c||c|c|c|c||c|c|c|c|}  \hline
  % after \\: \hline or \cline{col1-col2} \cline{col3-col4} ..
    area & $j_u$ & $j_{u+d}$ & $j_d$ & area & $j_u$ & $j_{u+d}$ & $j_d$ & area & $j_u$ & $j_{u+d}$ & $j_d$  \\ \hline
  $a_\mathrm{min}$ & 0 & $\frac{1}{2}$ & $\frac{1^{}}{2_{}}$ & $a_4$ & 0 & 1 & 1 & $a_6$ & $\frac{1}{2}$ & 0 & $\frac{1}{2}$
  \\ \hline
  $a_\mathrm{min}$ & $\frac{1}{2}$ & $\frac{1^{}}{2_{}}$ & 0 & $a_4$ & 1 & 1 & 0 & $a_6$ & $\frac{1}{2}$ & 2 & $\frac{3}{2}$
  \\ \hline
  $a_2$ & $\frac{1}{2}$ & 1 & $\frac{1^{}}{2_{}}$ & $a_4$ & 1 & 2 & 1 & $a_6$ & $\frac{3}{2}$ & 2 & $\frac{1}{2}$
  \\ \hline
  $a_3$ & $\frac{1}{2}$ & $\frac{3^{}}{2}$ & 1 & $a_5$ & 1 & $\frac{5}{2_{}}$ & $\frac{3}{2}$ & $a_6$ & $\frac{3}{2}$ & 3 & $\frac{3}{2}$
  \\ \hline
  $a_3$ & 1 & $\frac{3}{2}$ & $\frac{3^{}}{2}$ & $a_5$ & $\frac{3}{2}$ & $\frac{5}{2_{}}$ & 1 &  &  &  &  \\
  \hline
\end{tabular}
  \caption{The eigenstates corresponding to the first six
  SU(2)-valued spin networks.}\label{tab_su2}
\end{table}

The minimal area in this group is degenerate in the two states each
with a crossing spin one-half edge bending at the surface. In these
states there must be at least one completely tangential edges of
spin one-half connecting to the intersecting point.

Note that in the degenerate eigenvalues we investigated here, which
are the first a hundred levels, there is only one state that
exceptionally does not appear as a degenerate state and that is the
state with area $a_2$ in SU(2) group. In this state the upper and
lower edges of spins one-half connect at a vertex and bend at the
surface along the same direction.

\begin{figure}[h]
\begin{center}
\includegraphics[width=9cm]{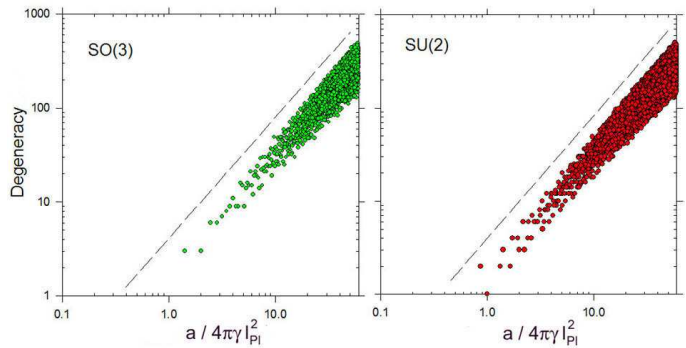}%\\(a)\\
\end{center}
\caption{The scatterplot of correlation between the area eigenvalues
and the degeneracies of their corresponding eigenstates.}
\label{cap. scatterplot 1}
\end{figure}

We counted the number of the degeneracy $g$ for different
eigenvalues in scatterplots and the area and its degeneracy appeared
to be correlated in Figure \ref{cap. scatterplot 1}. These
scatterplots indicates the results in SU(2) and SO(3) group
representations of spin networks separately in the  log-log graphs.

The eigenvalue degeneracy grows roughly as a power-law but with
increasing scatter, namely
\begin{equation}\label{eq. g(a)}
g(a) \approx  \left(\frac{a}{a_o}\right)^{\alpha + \epsilon},
\end{equation}
where $\alpha=0.96$ with the scatter uncertainty $\epsilon = \pm
0.03$. This particular degeneracy behaves scale invariant and is
robust, however we will so in this note later on that this is not
the only degeneracy of eigenvalues.

\begin{figure}[h]
\begin{center}
\includegraphics[width=9cm]{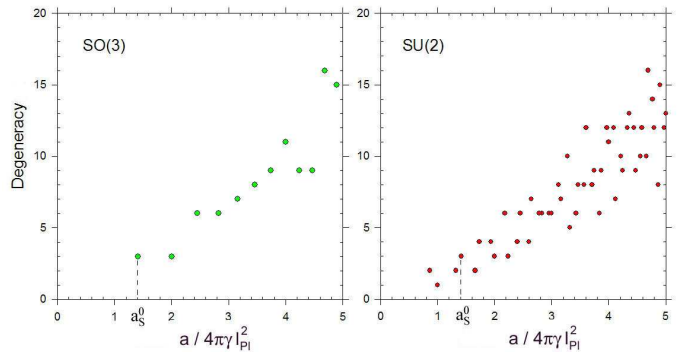}%\\(a)\\
\end{center}
\caption{The scatterplot of a few first area level degeneracies.}
\label{cap. scatterplot 2}
\end{figure}

For the purpose of further clarifications, we plot the minimal area
levels and their corresponding degeneracy separately in Figure
(\ref{cap. scatterplot 2}).

In fact it is obvious that we do not need two different plots for
the SO(3) and SU(2) groups because the spectrum of area in the SO(3)
group is contained in the SU(2) group degeneracy graph. For
instance, in Figure (\ref{cap. scatterplot 2}) we named the minimal
area in the SO(3) group by $a_o^S$ and showed this area is the
fourth level in the SU(2) group. More importantly this is the reason
why the exponent of the power law (\ref{eq. g(a)}) in the both group
representation is the same.

%\subsection{Degeneracy in generations}

Recently the complete spectrum of area is shown to be the union of
equidistant spectra, \cite{Ansari:2006vg}. Each one of the subsets
possesses a gap between levels equal to $a_o \chi \sqrt{\zeta}$. In
SO(3) group $\zeta$ is any square-free number $\{1,2,3,5,\cdots\}$
and the group characteristic parameter is $\chi := \sqrt{2}$; and in
SU(2) group $\zeta$ is the discriminant of any positive definite
form $\{3,4,7,8,11,\cdots\}$ and the group characteristic parameter
is $\chi := 1/2$. In other words, the complete spectrum reformulates
into $a_n(\zeta) = a_0 \chi \sqrt{\zeta}  n$ for $\forall\ n \in
\mathbb{N}$. Fixing $\zeta$ a generation of evenly spaced numbers is
singled out. The parameter $\zeta$ is therefore called the
generational number.

\begin{figure}[h]
  \includegraphics[width=9cm]{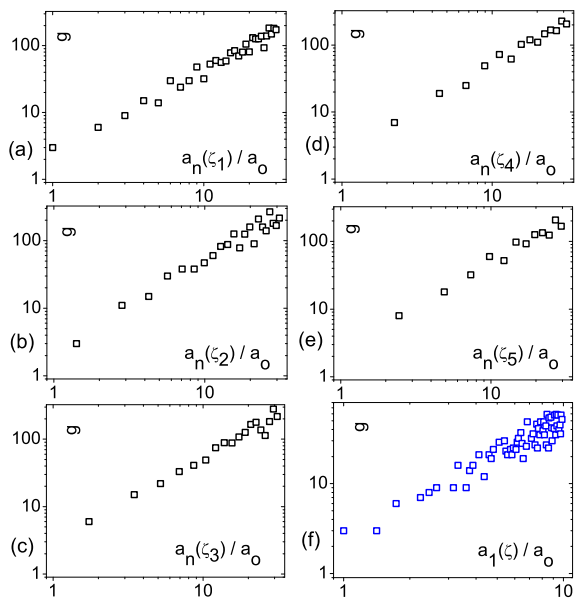}\\
  \caption{The generic degeneracy of figure (\ref{cap. scatterplot 1}) classified into generation in SO(3) group case. }\label{fig.generationdeg}
\end{figure}

Based on this classification, one can re-classify the generic
degeneracy of area levels into the generations. Plots (a-e) in
Figure (\ref{fig.generationdeg}) indicates this classification in
the first five generation. Plot (f) compares the degeneracy of the
first level of different generations.

Note that in any generation the area of higher levels can be
decomposed precisely into smaller fractions of the same generation
(without approximation) namely, $a_n = n a_1 = (n-2) a_1 + a_2 =
\cdots$. Let us consider for instance the configuration $n a_1$.
Each one of these area cells is degenerate. However, the states
corresponding to the eigenvalue in different regions of the surface
are distinguishable because by definition various number of
completely tangential edges with various spins could be connected to
each vertex without changing the area and the geometric
configurations. Therefore, the degeneracy of the area eigenvalue
$a_n$ is in fact $\Omega_n = g_n + g_{n-1}g_1 + \cdots + (g_1)^n$.
Obviously the dominant term in the sum belongs to the configuration
with maximum number of the area cell $a_1$.

\section{Entropy}

Consider a surface $S$ of a large area $A$. This area is the sum of
quanta in different configurations. In the dominant configuration it
contains the maximum number (N) of minimal degenerate area cell; $A
\approx N \amin$, where $N \gg 1$. Let us now consider the surface
is a black hole horizon. By the use of Einstein equation for a
collapsing star from a non-spherical state all radiatable
perturbations to the surface is radiated away such that at the late
stage the hole is left only with its monopoles. Now imagine the
initial deformation is located in a certain region of the horizon.
The future evolution of the field from that point depends on the
exact spin network state of that location, which includes the states
of completely tangential excitations. The state evolves under the
action of a Hamiltonian and is expected to radiate away energy from
the event horizon. Therefore, in the quantum states of a black hole
the complete information of spin network states makes regions
distinguishable from each other, although they may appear with the
same area. Having defined the dominant configuration for the
surface, the degeneracy of this configuration is therefore
$\Omega(A) = g(\amin)^N$. Consequently, the dominant entropy
associated to the underlying surface is proportional to $N \ln
g(\amin)$ or equivalently

\beq \label{eq. S} S = A\ \ln g(\amin) /\amin. \eeq

%\section{Black hole}

%\subsection{Bekenstein-Hawking entropy}

It is instructive to compare this result with the method the same
entropy is derived from the boundary picture. In that picture the
hole separates the space manifold into two sections, namely a
horizon boundary and the outer space. Gauge degrees of freedom are
still considered to be redundant in the bulk states but they become
physical degrees on the boundary. The reason is that the kinematical
Hilbert space of space includes the Hilbert spaces of the horizon
$\mathcal{H}_s$ and the bulk $\mathcal{H}_b$. The reduction of gauge
degrees of freedom  from this space takes place only in the bulk
partition because $\{\mathcal{H}_s \otimes \mathcal{H}_b \}/ SU(2)
\sim \mathcal{H}_s \otimes \{\mathcal{H}_b / SU(2)\}$. This means
the horizon Hilbert space accepts the gauge transformation
redundancies as the physical states. An edge of spin $j$ puncturing
the boundary produces $2j+1$ physical states on the surface. By
assuming that different area cells on the horizon are
distinguishable the dominant configuration is the one with a maximum
number of the minimal area $a_{\mathrm{min}}^{\mathrm{(punc)}}$ (the
area of spin $j_{\mathrm{min}}$ puncture). Consequently, black hole
entropy becomes $S = A \ln
(2j_{\mathrm{min}}+1)/a_{\mathrm{min}}^{\mathrm{(punc)}}$. Note
that, as it was mentioned above, the minimal quantum of area in this
calculation is not the minimal area in the complete spectrum of
area. Now the question is that based on what reason this subset was
first considered as the basis for representing the geometry of a
horizon? The easy answer to this is the one has been mentioned by
Rovelli in his original work \cite{carloblackhole} that none of the
other eigenvalues was known at the time the black hole entropy was
studied. However, there is a more sophisticated and physical answer
to this. As soon as one assumes a black hole horizon as the internal
boundary of space, the contribution of the boundary to the gravity
action of the space becomes the Chern-Simons action, \footnote{This
is also true in any BF theory action. More precisely nothing enters
into this surface term from the special character of gravity as a
constrained BF theory.}. When this space is quantized the geometry
of this boundary, in principle, is described by a set (and not a
sequence) of punctures. As it was mentioned above, the gauge degrees
of freedom on this boundary become physical. Regardless of the group
representation that suits the boundary fields (which could be either
SU(2) or U(1)), since the internal spin network states in this
approach are removed out, the boundary is left only with a subset
quanta (those which puncture the boundary).

However, beyond postulations and assumptions there is no physical
reason to outlaw considering the spin network states in the interior
region of black hole. Indeed, the black hole interior may be in an
infinite number of states. For instance, the black hole interior may
be given by a Kruskal extension so that on the other side of the
hole there is another universe. The inclusion of these states in
fact allows the horizon to be quantized via the complete spectrum of
area. However, the number of those internal states cannot affect the
interaction of the hole with its surroundings. From the exterior,
the hole is completely determined by the properties of its surface.
Thus, the entropy relevant for the thermodynamical description of
the thermal interaction of the hole with its surroundings is
determined by the states of the quantum gravitational field on the
black hole surface.

In principle, depending on the Hamiltonian operator that maps the
physical Hilbert space of the interior region of a closed surface
into itself, the surface may evolve and its area may increase,
decrease or become unchanged in the course of time. In the case of
black hole, the time evolution map is responsible for its
non-decreasing area character and turning the kinematic entropy we
reported here into a physical non-decreasing entropy,
\cite{Sorkin:2005qx}. This motivates a picture of a more quantum
black hole in which the quantization of space occurs prior to the
definition of the black hole sector. Here we restrict the results
into only the kinematical ones and disregard arguing about the
dynamics. Of course one way to start the definition of such a
horizon perhaps is possible through a spin foam model, (for review
of different models see \cite{Perez:2003vx}).

A non-rotating spherical black hole `responses' any perturbation by
some complex frequencies called \emph{quasinormal modes}. This makes
a black hole horizon state different from a random surface. The
imaginary part of the modes is the frequency at which the
perturbation is damped. The real part is the frequency at which a
quantum of energy is emitted from the black hole or is absorbed into
it. When the perturbation is damped very quickly, the response of
the black hole are the emission of the quantum of energy $\Delta M =
M^2_{\mathrm{Pl}} \ln 3/ 8 \pi  M$, \cite{Nollert:1999ji}. Since the
quanta of energy and area are proportional to each other $\Delta M =
\Delta A\ (c^4 /32 \pi G^2 M)$, the exchanging area from the black
hole is $\Delta A = 4 \ln 3\ \ell^2_{\mathrm{P}}$.

Following Dreyer's conjecture \cite{Dreyer:2002vy} for identifying
the classical perturbation of a black hole horizon in highly damping
mode with the transition between the two natural configurations, it
can be assumed $\Delta A = a_\mathrm{min}$. From this equivalence
the entropy of a black hole in Planck's area unit becomes

\beq \label{eq. bh entropy} S = A \ln g_1(\zmin)/ 4 \ln 3
 \eeq in \emph{any} group representation.

Unexpectedly in SO(3) group, since the degeneracy of the minimal
area is three, the two logarithms from the numerator and denominator
of (\ref{eq. bh entropy}) are canceled out and the
Bekenstein-Hawking entropy is verified. This lets us to calculate
the Barbero-Immirzi parameter by the size of minimal area. This
minimal area in this picture is a half times smaller than the
minimal area on a boundary. The reason is that here we do not
restrict the horizon to be a boundary of space and the internal spin
network exists and connects to the horizon surface. The
Barbero-Immirzi becomes tuned here to the value $\gamma = \frac{\ln
3}{\pi \sqrt{2}}$. This value is the double of what has been
reported using Dreyer's result in a boundary picture of horizon.

In summary, we reported that the complete spectrum of area possesses
eigenvalue degeneracy. This degeneracy with respect to area in both
group representations is power law with increasing scatter. However
since the complete spectrum of area is the union of different
equidistant subsets, the total degeneracy of a large eigenvalue
becomes proportional exponentially to its area. The black hole
entropy relevant for the thermodynamical interaction of a black hole
with its exterior region is the number of the quantum microstates of
the horizon which are distinguishable from the exterior of the hole.
We have obtained that the entropy is proportional to the area.
Moreover we derived the exact form of the Bekenstein-Hawking formula
when the minimal area is considered to be the quantum of area
emitted from the hole in its highly damping mode.

\section{Acknowledgement}

Helpful discussions with K. Krasnov, C. Rovelli, L. Smolin, and R.
Sorkin are acknowledged.

\end{document}